\documentclass[aps,prb,showpacs,twocolumn]{revtex4}
\usepackage{graphicx}
\usepackage{amsmath}
\usepackage{amsfonts}
\usepackage{color}

\begin{document}
\title{Phenomenology of Andreev reflection from first-principles transport theory}

\author{Awadhesh Narayan, Ivan Rungger, and Stefano Sanvito}
\affiliation{School of Physics, AMBER and CRANN Institute, Trinity College, Dublin 2, Ireland}

\date{\today}

\begin{abstract}
We study Andreev reflection in normal metal-superconductor junctions by using an extended 
Blonder-Tinkham-Klapwijk model combined with transport calculations based on density functional 
theory. Starting from a parameter-free description of the underlying electronic structure, 
we perform a detailed investigation of normal metal-superconductor junctions, as the separation
between the superconductor and the normal metal is varied. The results are interpreted by means 
of transverse momentum resolved calculations, which allow us to examine the contributions arising 
from different regions of the Brillouin zone. Furthermore we investigate the effect of a voltage bias 
on the normal metal-superconductor conductance spectra. Finally, we consider Andreev reflection 
in carbon nanotubes sandwiched between normal and superconducting electrodes.
\end{abstract}

\maketitle

\section{Introduction}

An electron incident on a superconductor from a normal metal, with an energy smaller 
than the superconducting energy gap, cannot propagate into the superconductor and 
thus should be perfectly reflected. However, Andreev discovered a mechanism for transmission,
in which an electron may form a Cooper pair with another electron and be transmitted across 
the superconductor. As a consequence of charge conservation a hole must be left behind, 
which, as a result of momentum conservation, should propagate in a direction opposite to 
that of the incident electron. This process is termed Andreev reflection~\cite{andreev00}. 

Apart from providing a confirmation for the existence of Cooper pairs and superconductor 
energy gaps~\cite{tinkham}, this process may also have applications in spintronics. It has 
been suggested that point contact Andreev reflection can be used to probe spin polarization 
of ferromagnets by fabricating ferromagnet-superconductor 
nanojunctions~\cite{buhrman-ferro,coey-ferro}. Materials-specific modelling of such experiments, 
however, is complex and so far it has been somehow unsatisfactory. For instance tight-binding 
based scattering theory~\cite{lambert-pb} and Green's functions theory~\cite{turek-pb} calculations 
found poor fits to the experimental data for ferromagnet-superconductor junctions, while 
produced excellent fitting to normal metal-superconductor junctions results. Based on this 
observation, Xia and co-workers suggested that there may be an interaction between the 
ferromagnet and superconductor which is not accounted for in the Blonder-Tinkham-Klapwijk 
(BTK) model~\cite{turek-pb}. Consequently, the simple interpretation and two-parameter BTK 
model fitting of experimental data to extract the spin polarization of various ferromagnets, was 
also called into question. More recently, Chen, Tesanovic and Chien proposed a unified model 
for Andreev reflection at a ferromagnet-superconductor interface~\cite{chien-model}. This is 
based on a partially polarized current, where the Andreev reflection is limited by minority states 
and the excess majority carriers provide an evanescent contribution. However, this model has 
also been called into doubt by Eschrig and co-workers~\cite{zutic-comment}. In particular, they 
pointed out that the additional evanescent component is introduced in an \textit{ad-hoc} manner, 
and that the resulting wavefunction violates charge conservation. So, the debate about the correct 
model to describe Andreev reflection at a ferromagnet-superconductor junction seems far from 
being settled.

Among other mesoscopic systems, Andreev reflection has also been measured in carbon 
nanotubes (CNTs)~\cite{dai-cnt}. There has been a theoretical study of normal 
metal-molecule-superconductor junction from density functional theory based transport 
calculations~\cite{lambert-fano}. In this study it was shown that the presence of side groups 
in the molecule can lead to Fano resonances in Andreev reflection spectra. Topological insulators, 
a very recent and exciting development in condensed matter physics, have also been shown to 
be characterized by perfect Andreev reflection~\cite{du-inas,cooperpair,HgCdTe,sanvito-andreev}.

Wang and co-authors have recently suggested performing a self-consistent calculation of the 
scattering potential to study Andreev reflection at normal metal-superconductor 
junctions~\cite{wang-selfcons}. They calculated the conductance for carbon chains sandwiched 
between a normal and a superconducting Al electrode and found different values depending on
whether or not the calculation was carried out self-consistent over the Hartree and 
exchange-correlation potential. However, the theoretical justification for such a self-consistent 
procedure is at present not clear. In particular, it is difficult to argue that the variational principle, 
which underpins the Hohenberg-Kohn theorems, is still obeyed when a pairing energy is added 
\textit{by hand} to the Kohn-Sham potential. In principle a rigorous self-consistent treatment should 
use the superconducting version of density functional theory~\cite{kohn-scdft}, which probably remains
computationally too expensive for calculating the interfaces needed to address a scattering problem. 
Given such theoretical landscape and the fact that a non self-consistent approach to density functional 
theory based transport calculations has shown excellent agreement to experimental results for normal 
metal-superconductor junctions, we follow this methodology in the present work.

In this paper, we study Andreev reflection in normal-superconductor junctions, including 
all-metal junctions and carbon nanotubes sandwiched between normal and superconducting 
electrodes. We take into account the atomistic details of the junction by using density functional 
theory to obtain the underlying electronic structure, and then employ an extended BTK model 
to solve the normal-superconductor scattering problem. Our transverse momentum resolved 
calculations allow identifying the contributions to conductance from different parts of the 
Brillouin zone. We also study the variation of conductance as a function of an applied potential 
difference between the electrodes for various normal metal-superconductor junctions, by 
performing approximate finite bias calculations.

After this introduction, the rest of our paper is organized as follows: in Section~\ref{formulation} 
we summarize the extended BTK model and Beenakker's formula, which we employ in this 
work. In the subsequent Section~\ref{results}, we present our results for Cu-Pb, Co-Pb and Au-Al junctions, 
as well as Al-CNT-Al junctions. We also include the computational details in each of these subsections. Finally, we conclude and summarize our findings in Section~\ref{conclusions}.

\section{Formulation}\label{formulation}

For the sake of completeness, here we briefly summarize the extended BTK 
model~\cite{btk-blonder} that we use to study Andreev reflection at a normal 
metal-superconductor interface. Following Refs.~[\onlinecite{beenakker2,turek-pb}], 
we begin with the Bogoliubov-de Gennes equation
\begin{equation}
 \begin{pmatrix} 
H_{\sigma} & \Delta e^{i\phi}\\
\Delta^{\ast}e^{-i\phi} & -H^{\ast}_{-\sigma}
\end{pmatrix}
\begin{pmatrix}
\psi_{e\sigma}\\ \psi_{h-\sigma}
\end{pmatrix}=
\varepsilon
\begin{pmatrix}
\psi_{e\sigma}\\ \psi_{h-\sigma}
\end{pmatrix}\;,
\end{equation}
\noindent where $H_{\sigma}$ is the single particle Hamiltonian for majority 
($\sigma=1$) and minority ($\sigma=-1$) spins, $\Delta$ is the pairing potential 
and $\psi_{e}$ and $\psi_{h}$ are respectively the electron and hole wavefunctions. 
The energy $\varepsilon=E-E_\mathrm{F}$ sets the reference to the Fermi energy,
$E_\mathrm{F}$. We follow the approach of Beenakker consisting in inserting a 
layer of superconductor in its normal state between the metal-superconductor interface. 
This ensures that at the fictitious normal metal-superconductor interface the only 
scattering process is Andreev scattering. 

\begin{figure}[h]
\begin{center}
  \includegraphics[scale=0.7]{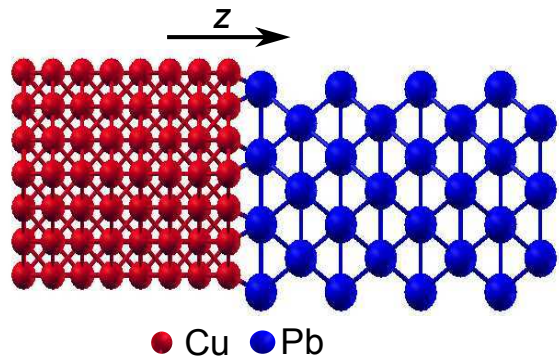}
  \caption{(Color online) Scattering region for a Cu-Pb junction. Transport calculations are 
  performed for different distances between Cu and Pb, employing periodic boundary
  conditions in the plane normal to the transport direction, $z$, indicated by the arrow. 
  Self-energies are used to simulate the effect of semi-infinite leads attached to the edge 
  of the scattering region.} \label{cu-pb-setup}
  \end{center}
\end{figure}
Other scattering processes are accounted for at the junction between the normal metal 
the and superconductor in its normal state. At this interface the scattering matrix can be
written as
\begin{equation}
 \begin{pmatrix}\psi_{1,e\sigma}^{-} \\ \psi_{2,e\sigma}^{+}\\ \psi_{1,h-\sigma}^{+}\\ \psi_{2,h-\sigma}^{-}\end{pmatrix}=\begin{pmatrix}s^{\sigma}(\varepsilon) & 0 \\0 & s^{-\sigma\ast} (-\varepsilon) \end{pmatrix}\begin{pmatrix}\psi_{1,e\sigma}^{+} \\ \psi_{2,e\sigma}^{-}\\ \psi_{1,h-\sigma}^{-}\\ \psi_{2,h-\sigma}^{+}\end{pmatrix}\:.
\end{equation}
Here the superscripts $+$ and $-$ denote the right- and left-going states and the subscripts 
$1$ and $2$ refer to the normal and fictitious normal metal regions, respectively. The normal 
state scattering matrix reads
\begin{equation}
 s^{\sigma}(\varepsilon)=\begin{pmatrix}r_{11}^{\sigma} & t_{12}^{\sigma} \\ t_{21}^{\sigma} & r_{22}^{\sigma}\end{pmatrix}\:.
\end{equation}

Now at the fictitious normal metal-superconductor interface 
\begin{equation}
 \psi_{2,e\sigma}^{-}=\alpha \psi_{2,h-\sigma}^{-}e^{i\phi}, \quad \psi_{2,h-\sigma}^{+}=\alpha^{\ast} \psi_{2,e\sigma}^{+}e^{-i\phi}\:,
\end{equation}
\noindent where the factor $\alpha$ is
\begin{eqnarray}
 \alpha &=& \exp[-i\cos^{-1}(\varepsilon/\Delta)], \quad |\varepsilon|<\Delta \nonumber \\
&=& \frac{1}{\Delta}[\varepsilon-\mathrm{sign}(\varepsilon)\sqrt{\varepsilon^{2}-\Delta^{2}}], \quad |\varepsilon|>\Delta\:.
\end{eqnarray}
\noindent The states in the normal metal are given by
\begin{equation}
 \begin{pmatrix}\psi_{1,e\sigma}^{-}\\ \psi_{1,h-\sigma}^{+}\end{pmatrix}= \begin{pmatrix}R_{ee}^{\sigma} & R_{eh}^{\sigma} \\ R_{he}^{\sigma} & R_{hh}^{\sigma}\end{pmatrix} \begin{pmatrix}\psi_{1,e\sigma}^{+}\\ \psi_{1,h-\sigma}^{-}\end{pmatrix}\:.
\end{equation}

\noindent Then the reflection coefficients for the complete system are
\begin{equation}\label{ree}
 R_{ee}^{\sigma}=r_{11}^{\sigma}(\varepsilon)+\alpha^{2}t_{12}^{\sigma}(\varepsilon)r_{22}^{-\sigma\ast}(-\varepsilon) \frac{1}{1-\alpha^{2}r_{22}^{\sigma}(\varepsilon)r_{22}^{-\sigma\ast}(-\varepsilon)} t_{21}^{\sigma}(\varepsilon)\:,
\end{equation}
and
\begin{equation}\label{rhe}
 R_{he}^{\sigma}=\alpha^{\ast}e^{-i\phi}t_{12}^{-\sigma\ast}(-\varepsilon) \frac{1}{1-\alpha^{2}r_{22}^{\sigma}(\varepsilon)r_{22}^{-\sigma\ast}(-\varepsilon)} t_{21}^{\sigma}(\varepsilon)\:.
\end{equation}

\noindent Finally the conductance of the system is given by
\begin{equation}\label{gnsfull}
 G_{NS}(\varepsilon)=\frac{e^{2}}{h}\sum_{\sigma=\pm 1}\mathrm{Tr}(1-R_{ee}^{\sigma}R_{ee}^{\sigma\dagger}+R_{he}^{\sigma}R_{he}^{\sigma\dagger})\:.
\end{equation}
The implicit assumptions in the above derivation are that the superconducting order 
parameter is switched on abruptly as a step function (i.e., there are no proximity 
effects) and the order parameter is much smaller than the Fermi energy (the so-called 
Andreev approximation). 

\begin{figure*}[ht]
\begin{center}
  \includegraphics[scale=0.65]{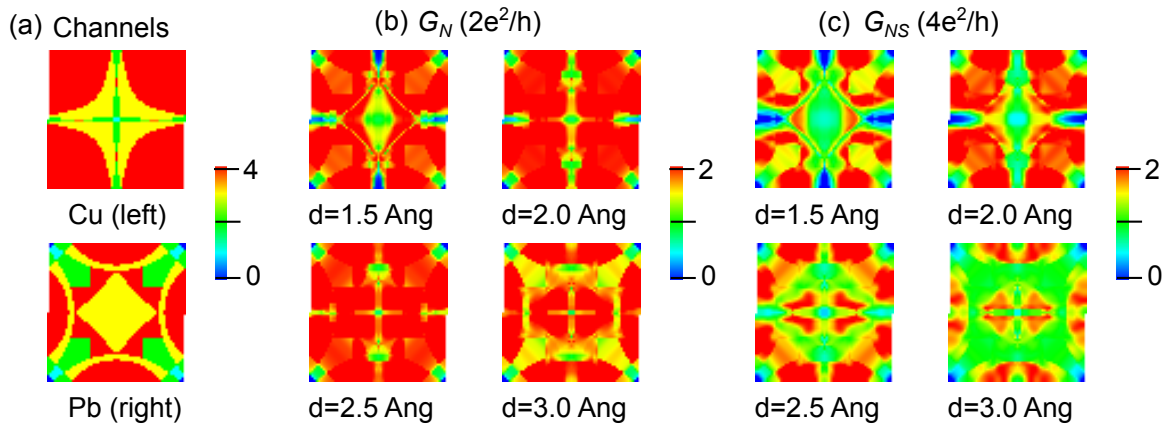}
  \caption{(Color online) Cu-Pb junction: $k$-resolved transport quantities at the Fermi level. 
  (a) A $k_{x}-k_{y}$ resolved plot of the channels in the left-hand (Cu) and right-right hand
  side (Pb) electrodes. (b) Normal conductance, $G_{N}$, and (c) normal 
  metal-superconductor conductance, $G_{NS}$, plotted across the Brillouin zone at different 
  Cu-Pb separations, $d$, ranging from 1.5 \AA{} to 3.0 \AA{}.}
  \label{cu-pb-kp}
  \end{center}
\end{figure*}

A great simplification occurs if one considers scattering at Fermi energy, namely $\varepsilon=0$, 
and the presence of time reversal symmetry, i.e., the normal metal is not a ferromagnet. The 
above expression for conductance reduces to 
\begin{equation}\label{gnsbeenakker}
 G_{NS}(\varepsilon=0)=\frac{4e^{2}}{h}\mathrm{Tr}\left(\frac{t_{12}t_{12}^{\dagger}}{2-t_{12}t_{12}^{\dagger}}\right)^{2}=\frac{4e^{2}}{h}\sum_{n}\frac{T_{n}^{2}}{(2-T_{n})^{2}}\:,
\end{equation}
where the eigenvalues of the transmission matrix product $t_{12}t_{12}^{\dagger}$ are 
$T_{n}$. This is the Beenakker's formula~\cite{beenakker1}. Notice that all the dependence 
on the superconductor pairing has dropped out and the conductance depends on the normal 
state transmission eigenvalues. In this case superconductivity enters implicitly in the form of 
a boundary condition. In our first-principles transport code 
{\sc Smeagol}~\cite{sanvito-smeagol1,sanvito-smeagol2,sanvito-smeagol3}, we construct the 
full scattering matrix and then use the expressions in equations~(\ref{ree}) and~(\ref{rhe}) to 
evaluate the conductance from equation~(\ref{gnsfull}). For the special case of 
$\varepsilon=E-E_\mathrm{F}=0$, we construct the transmission matrix, $t_{12}t_{12}^{\dagger}$. It is then straightforward to 
obtain its eigenvalues by numerical diagonalization. These are then interted into the 
Beenakker's formula [equation~(\ref{gnsbeenakker})] to obtain $G_{NS}$, while a direct 
summation of the eigenvalues yields $G_{N}=\frac{2e^{2}}{h}\sum_{n}T_{n}$. To compute 
the current, $I$, at a bias $V$, we use $G_{NS}$ from equation~(\ref{gnsfull}) and calculate
\begin{equation}
 I(V)=\frac{1}{e}\int d\varepsilon [f(\varepsilon)-f(\varepsilon+eV)]G(\varepsilon)\:,
\end{equation}
and the finite bias conductance is evaluated from
\begin{equation}
 G(V)=\int d\varepsilon\left(-\frac{\partial f}{\partial \varepsilon}\right)G(\varepsilon)\:.
\end{equation}
Here $G$ can either be the normal state conductance, $G_{N}$, or the normal 
metal-superconductor conductance, $G_{NS}$, and $f$ is the Fermi function.

\section{Results}\label{results}

\subsection{Cu-Pb junction}

We begin by presenting our results for Cu-Pb junctions, which have also been 
investigated experimentally in the past~\cite{buhrman-ferro}. We choose Cu $3d$ 
and $4s$ and Pb $6s$ and $6p$ as valence electrons and the effect of other core 
electrons are described by Troullier-Martins norm-conserving pseudopotentials. 
The local density approximation with the Ceperley-Alder parametrization was employed 
for the exchange-correlation functional. We choose an energy cutoff of 400 Rydberg 
for the real space mesh, and a double-$\zeta$ polarized basis set. The lattice constants 
of Cu ($a=3.615$\AA{}) and Pb ($a=4.93$\AA{}) are quite different, however a matching 
is obtained by rotating Cu unit cell by a $45^{\circ}$ angle. In this geometry a small strain 
($\approx 2$\%) exists on both Cu (compressive) and Pb (tensile). For the self-consistent 
calculation we use a $4\times 4$ in plane Monkhorst-Pack grid, while transport quantities 
are evaluated over a much denser $60\times 60$ $k$ grid. 

\begin{figure}[ht]
\begin{center}
  \includegraphics[scale=0.5,clip=true]{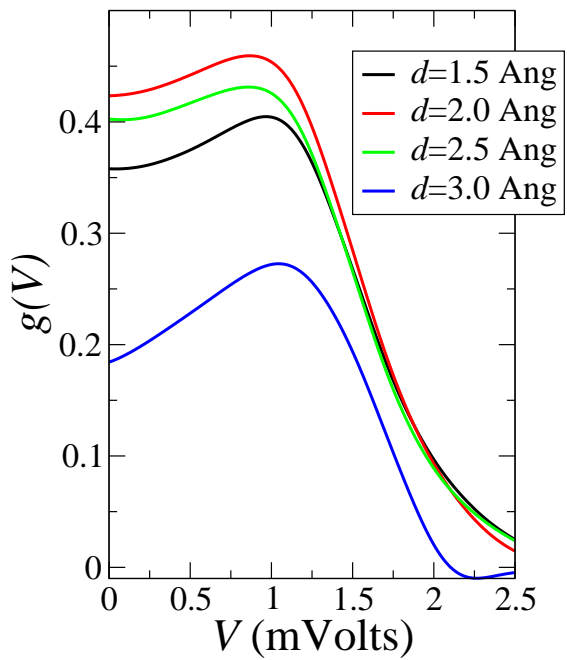}
  \caption{(Color online) Normalized conductance at finite bias for Cu-Pb junction at 
  different distances, $d$, between Cu and Pb. Note that $g(V=0)$ remains positive for 
  all the distances investigated here.}
  \label{cu-pb-bias}
  \end{center}
\end{figure}
\begin{figure}[h]
\begin{center}
  \includegraphics[scale=0.8]{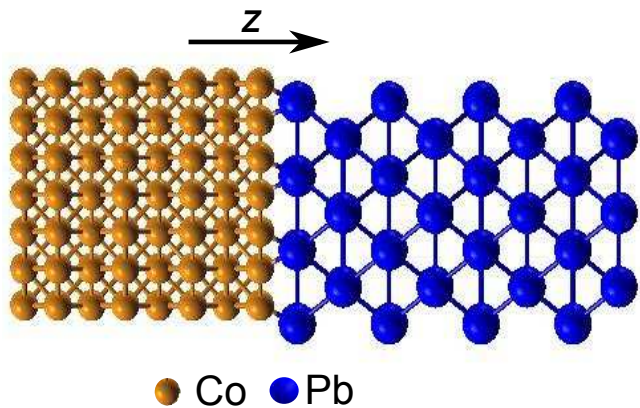}
  \caption{(Color online) Scattering region for the Co-Pb junction. Semi-infinite Co and Pb leads 
  are attached to the left-hand and right-hand side ends of the junction, respectively. 
  In the $x-y$ plane (orthogonal to transport direction, $z$) periodic boundary conditions are employed.}
  \label{co-pb-setup}
  \end{center}
\end{figure}

The scattering region for a Cu-Pb junction is shown in Fig.~\ref{cu-pb-setup}. We use 
periodic boundary conditions in the plane orthogonal to the transport direction, and 
open boundary conditions along the direction of transport. We plot the available channels 
for both electrodes resolved over the Brillouin zone (BZ) at the Fermi energy in 
Fig.~\ref{cu-pb-kp}(a). For the left electrode (Cu) four channels are available in 
quadrants centered at the edge of the BZ, with a residual region around the zone 
center in which either three or two channels are available. For the right electrode (Pb) 
around the BZ center there exists a rectangular region with three open channels, while 
at the BZ corners there are small pockets of reduced available channels, which even 
drop down to zero. The normal conductance, $G_{N}$, is large over almost the entire 
BZ, along with small pockets of lower transmission at the edges of the BZ, which are 
inherited from the reduced channel pockets in the Pb electrode, as shown in 
Fig.~\ref{cu-pb-kp}(b). Another small conductance pocket is present at the zone center, 
which originates from the distribution of open channels across the BZ in the Cu electrode. 

The overall conductance remains largely unchanged as the Cu-Pb distance is increased 
from $d=$~1.5 to 3.0 \AA{}. Next we show the normal metal-superconductor junction conductance, 
$G_{NS}$, in Fig.~\ref{cu-pb-kp}(c). At $d=$ 1.5 \AA{}, the pockets of small conductance 
at the zone edges are more prominent, as compared to $G_{N}$. Moreover, the region 
around $k_{x}=k_{y}=0$, with reduced conductance is also larger. On increasing the 
distance to 2 \AA{}, these low conductance pockets shrink in size and the overall conductance 
increases. At larger distances, a broader region of low conductance develops and this 
reduces overall $G_{NS}$. In Table~\ref{cu-pb-tab}, we provide the $k$-averaged value 
of the conductance above ($G_{N}$) and below ($G_{NS}$) the Pb superconducting 
temperature. For both quantities a maximum is obtained at $d=$ 2 \AA{}. We also tabulate 
the ratio $G_{NS}/G_{N}$, which is the quantity expressing the zero-bias suppression due to 
Andreev reflection. For a single channel BTK model describing an ideal interface this ratio is 
exactly two, however when one takes into account the band structure mismatch and the 
underlying electronic structure of the electrodes a much lower value for this ratio can be 
obtained. For the Cu-Pb equilibrium distance ($d\approx 2.0$ \AA{}) we find $G_{NS}/G_{N}$ 
close to 1.4, which is in excellent agreement with the experimental value of 1.38 reported in 
Ref.~\onlinecite{buhrman-ferro}. 

\begin{figure}[ht]
\begin{center}
  \includegraphics[scale=0.5,clip=true]{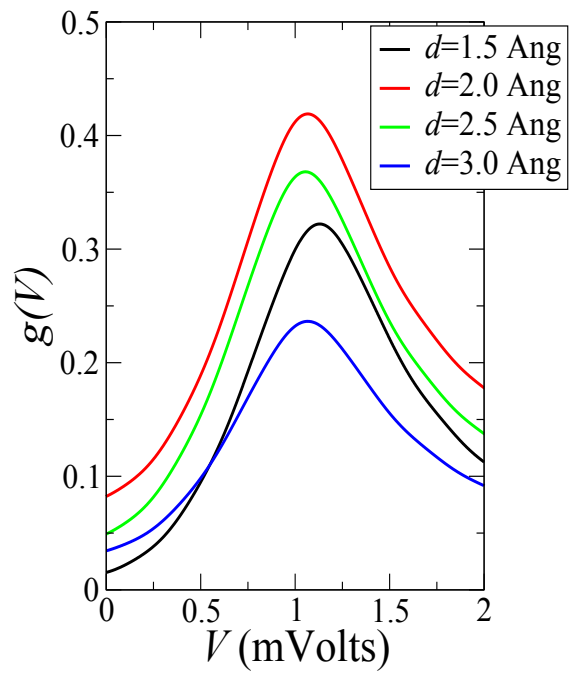}
  \caption{(Color online) Normalized conductance at finite bias for a Co-Pb junction at different 
  distances, $d$, between the two constituents.}
  \label{co-pb-bias}
  \end{center}
\end{figure}

\begin{table}[ht]
\caption{Cu-Pb junction: Normal conductance, $G_{N}$, normal-superconductor conductance, 
$G_{NS}$, and their ratio at different Cu/Pb distances.} \label{cu-pb-tab}
\centering 
\begin{tabular}{c c c c c} 
\hline \hline 
Distance (\AA{}) & $G_{N} (2e^{2}/h)$ & $G_{NS} (2e^{2}/h)$ & $G_{NS}/G_{N}$ \\[0.5ex]
\hline \hline
1.5 & 2.097 & 2.847 & 1.358\\
2.0 & 2.291 & 3.261 & 1.423\\
2.5 & 2.263 & 3.173 & 1.402\\
3.0 & 1.957 & 2.318 & 1.184\\ [1ex] 
\hline \hline
\end{tabular}
\end{table}

We also calculate the conductances at finite bias, for which we use the bulk Pb superconducting gap, 
$\Delta=1.36$ meV, and a temperature of 4.2 K in the Fermi distribution. The normalized conductance, 
$g(V)$, which is the quantity suitable for comparison across experiments, reads
\begin{equation}
 g(V)=\frac{G_{NS}(V)-G_{N}(V)}{G_{N}(0)}\:,
\end{equation}
where $G_{NS}(V)$ is the conductance at finite bias and temperature. Fig.~\ref{cu-pb-bias} shows $g(V)$ 
at different distances between the superconducting tip and the substrate. By large all the $g(V)$ curves follow 
a similar trend. They do start in the range [0.2, 0.4] for zero bias voltage, then they increase as the applied voltage 
gets larger. A maximum is achieved for $V$ close to the superconducting gap, and then $g(V)$ decreases for 
higher bias, since there is no Andreev reflection for energies above the superconducting gap. The normalized 
conductance curves for $d=1.5,2.0$ and $2.5$ \AA{} are quite closely spaced, while a considerable drop is seen 
for $d=3.0$ \AA{}. However, for small bias voltages of upto 2~mV, the normalized conductance remains positive, 
i.e., $G_{NS}>G_{N}$, for all distances to 3 \AA{}. This is consistent with experiments~\cite{buhrman-ferro}.

\subsection{Co-Pb junction}

Next, we present our results for Co-Pb junctions. We choose Co $3d$ and $4s$ and Pb $6s$ and $6p$ as valence 
electrons, while the core electrons are described by Troullier-Martins norm-conserving pseudopotentials. As in the 
Cu-Pb case, the local density approximation with the Ceperley-Alder parametrization was employed for the 
exchange-correlation functional and an energy cutoff of 400 Rydberg was chosen for the real space mesh. 
We selected a double-$\zeta$ polarized basis set. Similarly to the case of the Cu-Pb junction, we face the issue that the Co 
($a=3.548$\AA{}) and Pb ($a=4.93$\AA{}) lattice constants are quite different. Once again, we obtain a good lattice
match by rotating the Co unit cell by a $45^{\circ}$ angle, and the scattering region setup is shown in Fig.~\ref{co-pb-setup}. 
A $4\times 4$ in plane Monkhorst-Pack grid $k$-point sampling was used for the self-consistent calculation, and the transport 
quantities were evaluated over a dense $60\times 60$ $k$-grid. Note that Co is ferromagnetic and its spin polarization 
was determined based on a fit to the two parameter BTK model in Refs.~[\onlinecite{buhrman-ferro,coey-ferro}]. 

The zero bias conductances, $G_{N}$ and $G_{NS}$, along with their ratio is presented in Table~\ref{co-pb-tab}. 
There is a small variation of both $G_{N}$ and $G_{NS}$ as the substrate-superconducting tip distance, $d$, is changed. 
We note that the ratio $G_{NS}/G_{N}$ is reduced (to values close to one), in comparison to the Cu-Pb case. This is a 
manifestation of the ferromagnetism, where only one spin channel dominates around the Fermi level. As a consequence 
Andreev reflection is suppressed, since there are no opposite spin channels available for the hole and this hinders the 
formation of the Cooper pair in the superconductor. The values of this ratio obtained from our calculations are in good 
agreement with experimental studies~\cite{buhrman-ferro}, as well as previous calculations~\cite{lambert-pb,turek-pb}. 

The apparent match between experiments and theory is lost once one considers a finite bias situation. This can be appreciated
in Fig.~\ref{co-pb-bias}, where we plot the normalized conductance, $g(V)$, for different $d$. For all the distances investigated, 
$g(V)$ presents a similar shape. The curves start from a small value comprised between 0 and 0.1 around zero bias. 
They then increase and reach a maximum at around 1.3~mV. The main noticeable feature is the strong enhancement 
of $g(V)$, i.e., an increased Andreev reflection, at voltages close to the superconducting gap edge. Our results match 
previous calculations by Xia \textit{et al.}~\cite{turek-pb}, but they do not fit well the measurements of Upadhyay \textit{et al.} 
[\onlinecite{buhrman-ferro}]. In fact, in actual samples such strong enhancement of $g(V)$ near the superconductor 
gap edge is not seen. This discrepancy was attributed to the possibility of stray fields from the ferromagnet, which can 
cause spin-dependent splitting of the superconductor density of states~\cite{turek-pb}. Other possible reasons may include 
proximity effects, which are not taken into account in the extended BTK model and require a full DFT description of the 
superconducting state across the junction.

\begin{figure}[h]
\begin{center}
  \includegraphics[scale=0.8]{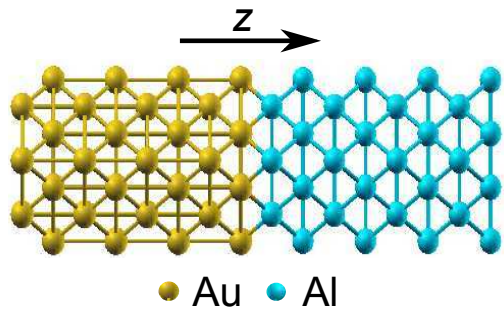}
  \caption{(Color online) Scattering region for an Au-Al junction. Au and Al semi-infinite leads 
  are attached to the left-hand and right-hand side ends of the junction, respectively. 
  Periodic boundary conditions are used in the plane perpendicular to the transport direction, $z$.}
  \label{au-al-setup}
  \end{center}
\end{figure}
\begin{figure*}[ht]
\begin{center}
  \includegraphics[scale=0.65]{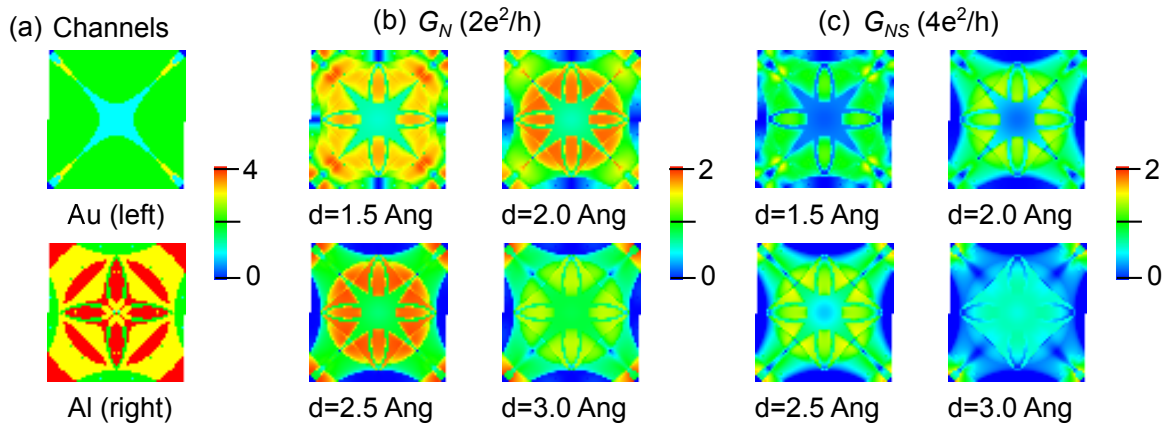}
  \caption{(Color online) Au-Al junction: $k$-resolved transport quantities at the Fermi level. 
  (a) A $k_{x}-k_{y}$ plot of the number of channels available in the left-hand side Au electrode 
  (top panel) and the right-hand side Al electrode (bottom panel). (b) Normal conductance 
  ($G_{N}$) across the junction plotted across the Brillouin zone orthogonal to the transport 
  direction at different Al/Au distances, $d$, in its normal state. (c) A $k_{x}-k_{y}$ resolved 
  plot of the normal metal-superconductor conductance ($G_{NS}$) at different $d$. Note the 
  same scale on the color plots for $G_{N}$ and $G_{NS}$, while different units are used for 
  the two conductances.}
  \label{au-al-kp}
  \end{center}
\end{figure*}

\begin{table}[ht]
\caption{Co-Pb junction: Normal conductance, $G_{N}$, normal-superconductor conductance, 
$G_{NS}$, and their ratio at different Co/Pb distances.} \label{co-pb-tab}
\centering 
\begin{tabular}{c c c c c} 
\hline \hline 
Distance (\AA{}) & $G_{N} (2e^{2}/h)$ & $G_{NS} (2e^{2}/h)$ & $G_{NS}/G_{N}$ \\[0.5ex]
\hline \hline
1.5 & 1.659 & 1.679 & 1.012\\
2.0 & 1.668 & 1.803 & 1.081\\
2.5 & 1.627 & 1.705 & 1.048\\
3.0 & 1.539 & 1.588 & 1.032\\ [1ex] 
\hline \hline
\end{tabular}
\end{table}

\subsection{Au-Al junction}

\begin{figure}[ht]
\begin{center}
  \includegraphics[scale=0.5,clip=true]{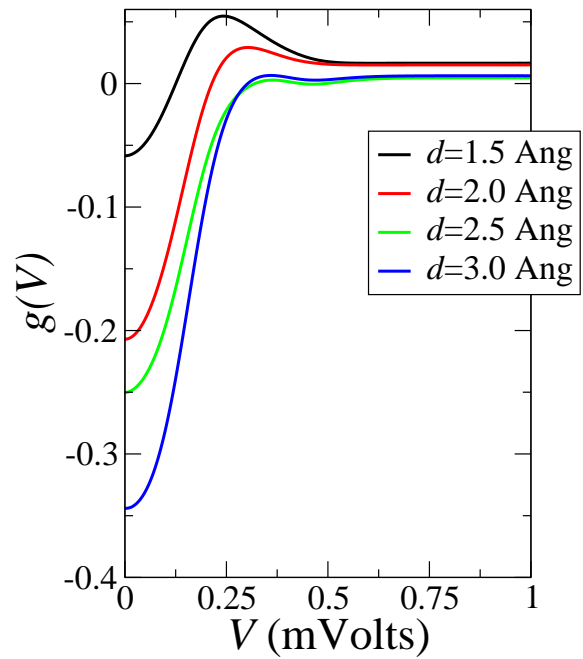}
  \caption{(Color online) Normalized conductance at finite bias for a Au-Al junction at different 
  Au/Al distances, $d$.}
  \label{au-al-bias}
  \end{center}
\end{figure}

Next, we move on to Au-Al junctions. We considered Au $5d$ and $6s$ and Al $3s$ and $3p$ 
as valence electrons and the core electrons are described again by Troullier-Martins norm-conserving
pseudopotentials. Similarly to the case of the Cu-Pb junction, the local density approximation with 
the Ceperley-Alder parametrization is employed for the exchange-correlation functional. We choose 
an energy cutoff of 400 Rydberg for the real space mesh, and a double-$\zeta$ polarized basis set. 
Since the lattice constants of Au ($a=4.078$\AA{}) and Al ($a=4.037$\AA{}) are quite close, a 
matching Au-Al junction is constructed with a small strain of $\approx 1$\% on Au. For the 
self-consistent calculation, we used a $4\times 4$ in plane Monkhorst-Pack grid. From this 
converged density, the various transport quantities are evaluated using a $60\times 60$ $k$-point 
grid.

\begin{figure}[h]
\begin{center}
  \includegraphics[scale=0.7]{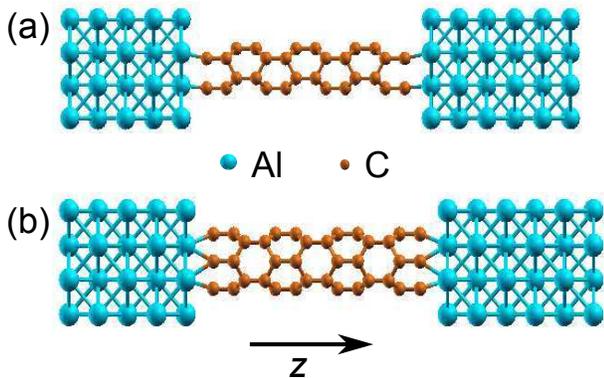}
  \caption{(Color online) Scattering region for (a) Al-CNT(3,0)-Al and (b) Al-CNT(4,0)-Al junctions. 
  In these two cases the same Al leads are attached on both sides of the scattering region.}
  \label{al-cnt-setup}
  \end{center}
\end{figure}

The scattering setup is shown in Fig.~\ref{au-al-setup}, with transport along the $z$ direction. 
Similarly to the Cu-Pb junction, we use periodic boundary conditions in the plane perpendicular 
to the transport direction and attach self-energies corresponding to semi-infinite Au and Al on the 
left-hand and right-hand sides, respectively. The $k$-resolved quantities are presented in 
Fig.~\ref{au-al-kp}. Over most of the BZ, two channels are available for the Au electrode, which 
reduces to a single channel around the zone center ($k_{x}=k_{y}=0$). The Al electrode, in contrast, 
provides a minimum of two channels over the entire BZ. The normal conductance $G_{N}$ 
(i.e. the conductance of the junction when the temperature is greater than the critical 
superconducting temperature for Al), is shown in Fig.~\ref{au-al-kp}(b) for different distances 
between Au and Al. 

The qualitative picture obtained for the four distances investigated in this work is similar. 
There is a four-fold symmetric feature at the zone center, which originates from the available 
states in the Au electrode. Around this central feature, a circular maximum in conductance is 
seen. This resembles the $k$ distribution of channels in the Al electrode. Its intensity  increases 
as one goes from a distance of 1.5 \AA{} to 2.5 \AA{}, but subsequently falls at $d=$3 \AA{}. 
In Fig.~\ref{au-al-kp}(c) we plot the normal metal-superconductor conductance $G_{NS}$ of 
the Au-Al junction. In this case, as for $G_{N}$, the four-fold symmetric feature arising from 
Au channels is seen around the BZ center. However, at $d=$2 and 2.5 \AA{}, a circular region 
with reduced conductance value develops around $k_{x}=k_{y}=0$. Furthermore, the circular 
feature in the normal conductance, obtained from the channels in Al electrode, takes a more 
dispersed shape in $G_{NS}$. Yet, the general trend that this region contributes most to the
conductance remains. The values of $G_{N}$ and $G_{NS}$ averaged over the entire BZ, 
for different distances, are summarized in Table~\ref{au-al-tab}. For all the four distances 
investigated, $G_{N}$ remains close to around one quantum of conductance, while it decreases 
as the distance between Au and Al is increased. Following a similar trend, $G_{NS}$ also 
decreases with increasing separation. In this case of a Au-Al junction the ratio $G_{NS}/G_{N}$ 
is obtained to be smaller than unity for all the four distances, in contrast to Cu-Pb junctions. 
This means that for Au-Al junctions the normal state conductance is larger than the superconducting 
state conductance. This provides a stark contrast to the expectation of doubling the conductance 
upon switching on superconductivity provided by the BTK model. Our results also emphasize 
the importance of taking into account the electronic structure of the materials forming the junction, 
as well as the Fermi surface mismatch between them.

\begin{table}[ht]
\caption{Au-Al junction: Normal conductance, $G_{N}$, normal-superconductor conductance, 
$G_{NS}$, and their ratio at different Au/Al distances.} \label{au-al-tab}
\centering 
\begin{tabular}{c c c c c} 
\hline \hline 
Distance (\AA{}) & $G_{N} (2e^{2}/h)$ & $G_{NS} (2e^{2}/h)$ & $G_{NS}/G_{N}$ \\[0.5ex]
\hline \hline
1.5 & 1.232 & 1.161 & 0.942\\
2.0 & 1.146 & 0.909 & 0.793\\
2.5 & 1.074 & 0.806 & 0.751\\
3.0 & 0.858 & 0.563 & 0.656\\ [1ex] 
\hline \hline
\end{tabular}
\end{table}

Finally in Fig.~\ref{au-al-bias} we plot the normalized conductance for different Au-Al 
separations. The superconducting gap for Al is $\Delta=0.17$ meV and we select a temperature 
of 1.2 K in the Fermi distribution. In contrast to the previous two junctions, in this case $g(V)$ is negative at low bias, since $G_{NS}$ is 
smaller than $G_{N}$. Furthermore, in this small bias regime increasing $d$ leads to a reduction 
in the value of the normalized conductance. For larger bias values, $g(V)$ rises as the voltage bias is increased and $G_{NS}$ and $G_{N}$ become 
almost equal and the normalized conductance tends to zero.

\subsection{Al-CNT-Al junctions}

We now discuss our investigation of Al-CNT-Al junctions. We select Al $3s$ and $3p$ and 
C $2s$ and $2p$ as valence electrons, and again norm-conserving pseudopotentials describe 
the core electrons. As before, we choose the local density approximation to the 
exchange-correlation functional, a real space mesh cutoff of 400 Rydberg and a double-$\zeta$ 
polarized basis set. We use a $4\times 4$ in plane $k$-point grid transverse to transport direction 
both for converging the charge density and for evaluating the conductances. 

\begin{figure}[ht]
\begin{center}
  \includegraphics[scale=0.5,clip=true]{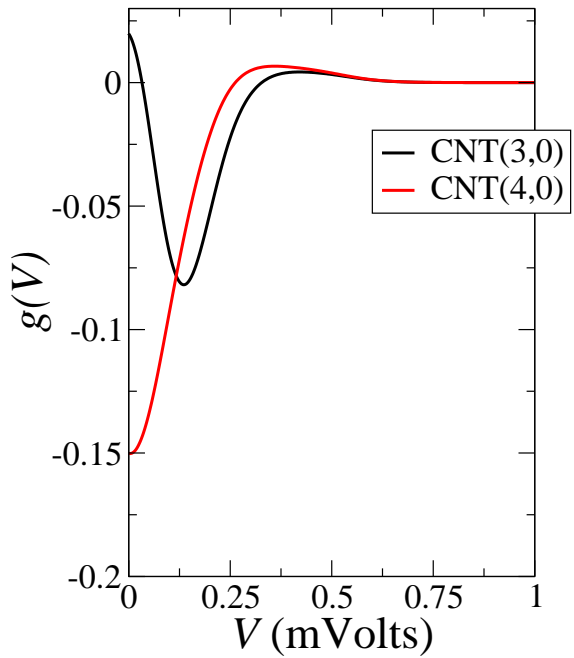}
  \caption{(Color online) Normalized conductance at finite bias for Al-CNT-Al junctions. 
  For CNT(3,0) $g(V)$ starts from a positive value and changes sign with the applied bias, 
  while for the case of CNT(4,0) it remains negative.}
  \label{al-cnt-bias}
  \end{center}
\end{figure}

We study $(3,0)$ and $(4,0)$ CNTs sandwiched between Al electrodes and we assume 
that one of them can be turned into a superconductor below a critical temperature. In practice 
this may be achieved by a proximity effect with a superconductor with a higher critical temperature, 
for example Sn, Pb or Nb. We fix the distance between the CNTs and the Al surface to 2 \AA{}, and
the CNT is left open-ended. The scattering region setup for the two cases is shown in 
Fig.~\ref{al-cnt-setup}. We have checked that the number of available states in Al leads is at 
least eight over the entire BZ. For Al-CNT$(3,0)$-Al junction, the major contribution to the 
normal conductance, $G_{N}$, comes from the zone center. A similar feature is seen for 
Al-CNT$(4,0)$-Al junction, where the conductance is dominated by the BZ center. The 
normal-superconductor conductance, $G_{NS}$, presents a very similar BZ picture for both 
types of CNTs. The $k$-averaged values are summarized in Table~\ref{al-cnt-tab}. The 
conductance values for the two CNTs are found to be similar. Interestingly, we find the ratio 
$G_{NS}/G_{N}$ to lie on opposite sides of unity; for $(3,0)$ CNT it is above one, while it falls 
below one for $(4,0)$ junction. This is reminiscent of the even-odd effect in C wires sandwiched 
between Al superconducting electrodes. Wang \textit{et al.} found that the zero bias Andreev conductance 
oscillates as the number of C atoms is changed from even to odd~\cite{wang-selfcons}.
\begin{table}[ht]
\caption{Al-CNT-Al junction: Normal conductance ($G_{N}$), normal-superconductor conductance ($G_{NS}$) and their ratio for different CNT's.} \label{al-cnt-tab}
\centering 
\begin{tabular}{c c c c c} 
\hline \hline 
CNT  & $G_{N} (2e^{2}/h)$ & $G_{NS} (2e^{2}/h)$ & $G_{NS}/G_{N}$ \\[0.5ex]
\hline \hline
(3,0) & 1.362 & 1.389 & 1.020\\
(4,0) & 1.431 & 1.217 & 0.850\\ [1ex] 
\hline \hline
\end{tabular}
\end{table}

Finally, we study the variation of normalized conductance as a function of an applied bias, 
which is plotted in Fig.~\ref{al-cnt-bias}. At low bias, for Al-CNT(3,0)-Al junction the normal 
metal-superconductor conductance is greater than the normal conductance. Interestingly, 
the situation is reversed at a voltage of 0.1~mV. In contrast, for Al-CNT(4,0)-Al junction, 
the normalized conductance remains negative for voltages less than the superconducting 
gap.

\section{Conclusions and future direction}\label{conclusions}

In conclusion, we have studied Andreev reflection in normal-superconductor junctions 
using density functional theory based transport calculations. This approach allowed us 
to include the atomistic details of the junction electronic structure in the extended
Blonder-Tinkham-Klapwijk model. We studied Au-Al and Cu-Pb all metal junctions and 
calculated the normal and normal-superconductor conductances for different separations 
of the two materials at the interface. Our transverse momentum resolved analysis has allowed 
us to identify contributions to these quantities from different parts of the Brillouin zone. 
We found that the conductances for junctions in the superconducting state follows a similar 
$k$-point dependence as the normal state conductance. In other words, Andreev reflection is 
higher in Brillouin zone regions, where transmission is also high. 

We have also investigated Co-Pb ferromagnet-superconductor junctions. In this case, while at 
zero bias, our results satisfactorily match the experimental reports, a discrepancy was revealed at 
a finite bias, particularly at voltages close to the superconductor gap. This could possibly be attributed 
to stray magnetic fields from the ferromagnet or to proximity effects, both causes which are not included 
in the extended Blonder-Tinkham-Klapwijk model. 

We further studied Andreev reflection from carbon nanotubes sandwiched between normal metal and 
superconducting electrodes and found $G_{NS}/G_{N}$ ratios to lie on opposite sides 
of unity for $(3,0)$ (higher than one) and $(4,0)$ (lesser than one) carbon nanotubes.
This highlights the sensitivity of such calculations to details and the need for a truly
atomistic theory for tackling this problem. 

Concerning the potential outlook for future studies, our work provides a stepping stone for analyzing with 
first-principles methods the experimental setups needed to investigate and detect Majorana fermions. 
These particles, which are their own anti-particles, are expected to play a crucial role in topological quantum 
computing and have recently garnered significant attention in the condensed matter community. After several 
theoretical proposals, signatures of this particle were found experimentally in large spin-orbit nanowires in proximity 
with superconductors~\cite{majorana1,majorana2}. However, a number of issues remain unresolved and 
important questions need to be answered to confirm that indeed Majoranas were observed. Our implementation 
of the phenomenology of Andreev reflection in a first-principles approach can be quite useful to study such a 
setup, in particular, by taking into account the underlying electronic structure of the nanowires. When combined 
with the order-$N$ implementation of our {\sc Smeagol} code~\cite{sanvito-Si-on}, which allows us treating thousands 
of atoms, it opens the opportunity of recreating theoretically the aforementioned experiments in an \textit{ab inito} 
manner, which till now have been modelled empirically.

\section*{Acknowledgments}

AN is financially supported by Irish Research Council under the EMBARK initiative. IR and 
SS acknowledge additional support by KAUST (ACRAB project). The computational resources 
have been provided by Trinity Centre for High Performance Computing.

\end{document}